# E-MRS Strasbourg (France)

*MANUSCRIPT COVER PAGE FORM*

**Corresponding Author**   :Seref Kalem

**Full Mailing Address**    : **TUBITAK-UEKAE Institute of Electronics PK:74 GEBZE 41470 KOCAELI, TURKEY**

**Telephone**    : **+90 262 648 1622**

**Fax**    : **+90 262 648 1100**

**E-mail**    : **s.kalem@uekae.tubitak.gov.tr**

**OPTICAL CHARACTERIZATION OF DISLOCATION FREE Ge and GeOI WAFERS**, S. Kalem, TUBITAK - Institute of Electronics, Gebze 41470 Kocaeli, Turkey. I. Romandic, A. Theuwis, UMICORE Electro-Optic Materials, 2250 Olen, Belgium.

**Abstract**

Optical properties of dislocations free Germanium and Germanium-on-insulator wafers have been characterized using Fourier transformed infrared spectroscopy, attenuated total reflectance, laser Raman scattering, linear and nonlinear optical transmission. In n-type Ge, localized vibrational modes of unintentional impurities are observable at 451, 488, 774 cm$^{-1}$, a broad absorption band at 1235 cm$^{-1}$ and a group of vibrations at around 1600, 2350 and 3750 cm$^{-1}$. In p-Ge, the latter features are rather weak in intensity and only an absorption dip at 808 cm$^{-1}$ is observable after annealing at 400C. The most important feature in ir transmission is the appearance of new bands at unpolarized oblique angle of incidence, which can be explained by the presence of birefringence effects in Ge. These results are compared with the intrinsic bulk Ge to determine the effect of doping on the chemical structure of the wafers. Angle dependent absorption was also observed in GeOI where disorder induced LO-TO coupling mode at 1250 cm$^{-1}$ can only be observed at non-normal incidence. Sub-gap optical absorption reveals the presence of higher electronic density of states in the p-type wafers, which is supported also by Ge-Ge phonon line broadening as evidenced from Raman scattering. Nonlinear optical absorption properties were determined at 10.6 μm using a $CO_2$ laser. The results indicate clear differences in nonlinear behavior of n-Ge, p-Ge and GeOI wafers.

**Introduction**

Germanium(Ge) has recently been receiving a great deal of attention for its promising applications in variety of fields. It has already been the best semiconductor of choice for producing solar cells to use in space applications. It is also used in high energy detection and infrared optics. Now, the use of Ge in nano electronics is being considered for its higher hole carrier mobilities compared to Silicon (Si). In addition, the existence of high dielectric constant (k) insulator with low interface trap densities offer possible realization of Ge based high speed devices. Moreover, the availibity of dislocations free Ge wafer enables industry to develop Ge-on-insulator (GeOI) substrates. Bonding of Ge on Si through an intermediate layer of $SiO_2$ offers other application possibilities. Optoelectronics circuits can be fabricated on Ge side while electronic ones are produced on Si wafer. For a successful application, Ge has to meet stringent requirements. One of them requires that Ge is free from extended lattice defects which depend on the presence and properties of point defects introduced during crystal pulling or wafer processing. Extrinsic point defects in Ge have been described elsewhere in greater detail[1]. Despite their importance, the properties of intrinsic and extrinsic point defects in Ge are not very well known. Therefore, efforts creating further knowledge on defects in Ge would be very valuable.

**Experimental**

This work reports the characterization of n type Sb doped Ge , p type Ga doped Ge and GeOI (n-type Ge bound to n-type Si through a 0.5 μm $SiO_2$ layer). The characterization used FTIR, ATR(Attenuated Total Reflectance), Raman light scattering and optical transmission at around the fundamental band gap. Ge wafers always contain some Carbon(from the graphite crucible) and Silicon(from the feedstock) impurities. Metallic impurities such as Fe, Cu, Co, Ni,... are below the detection limits of DLTS. If they are present, their concentration should be within the ppt or tens of ppt range.

Annealing experiments were carried out at 400C under vacuum for 30minutes. Savitzky-Golay smoothing has been used to remove the effect of the periodic fringes in infrared spectra. Raman scattering used 35mW He-Ne laser at 632.8 nm at room temperature. A tunable CO2 laser and UV-VIS broad band fiber light source were used in pump-probe transmission experiments at room temperature.

**Results and discussion**

Figure 1 shows typical FTIR spectra of an n-Ge and an undoped(intrinsic) Ge wafer at normal incidence using unpolarized light. In addition to absorption bands in i-Ge, we observe vibrations at 451, 488, 774, a broad band at 1230 cm-1 and strong features at around 1600, 2350 and 3750 cm-1. Angle of incidence was increased to benefit from the greater effect of effective thickness. The spectrum taken at $40^o$ oblique incidence exhibit new absorption peaks at 463, 497, 535 , 667 and 940 cm-1. The features at higher energies lose their strength at oblique incidence. Whereas, the band at 666cm-1 grows enormously in strength in detriment of other bands as evidenced at $40^o$ oblique incidence. The annealing at 400C introduces new bands at 463, 535 and 649cm-1. Moreover, the band at 666cm-1 dominates others again in oblique incidence. The band at 667cm-1 has been observed in n-Ge at 80K and attributed to an oxygen vacancy VO- [2]. There is no significant change in the strength of the bands at 1600 and 3750cm-1 upon annealing. The strength of these bands decrease with oblique incidence and that of the band 2350cm-1 almost vanishes. The appearance of strong absorption bands under oblique incidence of radiation was reported in cubic crystal films and attributed to longitudinal optic(LO) modes(Berreman effect)[3]. But, this effect has only been observed in the p-polarized component of the light. Whereas, we observe it with unpolarized light. The effect cannot be explained by an increase in optical pathway due to high refractive index of Ge(only 1.3% increase at $40^o$). It can be argued that the light is polarized during the passage in Ge. As evidenced from the ir spectra, the light make many reflections and during this multiple pass effect it is p-polarized in optical birefringent field of Ge. In general, free carriers in a semiconductor can exhibit birefringence if the effective mass is anisotropic [4].

In FTIR spectra of p-Ge (0.025 Ohm-cm, p=$10^{17}$cm$^{-3}$) at normal and oblique incidence are shown in Fig.2. At low energy side, there is a band structure due to LH-HH(Light Hole to heavy hole) sub-band transitions in the valence band at k≠0. The other features belong to LH-SO(Light hole-Spin Orbit spliting band) and HH-SO transitions. There is no difference between the oblique and normal incidence apart from the absence of interference fringes at 80°. Shown also in this figure is the spectrum of p-type Ge annealed at 400°C. There is not any significant change in the band structure except the appearance of a feature at 808cm$^{-1}$ as shown at the insert in Fig.2.

ATR gives information about the surface species in Ge. The spectra exhibit features at around 1600, 2300 and 3400cm-1 which are also observed in IR transmission in both n and p type Ge. Their strength is stronger in n-Ge. These bands are not removed upon annealing at 400C. Their energetic positions correspond to 198, 290 and 465meV. Therefore, we dont rule out the possibility of attributing these features to deep levels in Ge. Deep acceptor traps at Ev+0.3 eV and Ec-0.4 eV were reported in Ge[5].

Optical absorption around the fundamental absorption edge is shown in Fig.5. There is an increase in absorption in the sub-gap region(λ>1600nm) indicating probably the presence of an increased electronic defect density due to disorder in p-Ge. This is also supported by a broadening of Ge-Ge phonon line at 301.1cm-1 in Raman spectrum(Fig.4). Other feature is the red shift at the absorption edge, indicating a band gap narrowing by 10 meV. There is a difference in the measured band width at half maximum for the Ge-Ge band between n-Ge and p-Ge. The band width of p-Ge is 4.8cm-1, that is 14,3% greater than n-Ge. This is supportive of higher sub-gap absorption observed in p-Ge(Fig.4).

This absorption is due to transitions from defect states near band edge which are probably induced by disorder. The phonon line broadening in Raman spectrum can be attributed to a disorder induced scattering.

Both the n-type and p-type Ge wafers show nonlinear optical absorption effects. The results of transmission difference ( $\Delta T=T-T_0$) before($T_0$) and after(T) pump switched on are summarized in Fig.3. In p-Ge, the transmission is enhanced while it is decreased in n-Ge. At 10.6m, the CO2 laser is in resonance with sub-band transitions in p-Ge. The pump creates enough holes to fill in valence band states, thus enhancing transparency to CO2 laser. In n-Ge, the free carriers is probably absorbing the CO2 laser line. Thus, the transmission is decreased with the creation of extra free electrons by pumping.

FTIR is effective in mapping out interfaces, thus enabling us to assess the bonding quality in GeOI wafers. There are regions where the ir transmission spectrum is characteristic of a rough surface, thus indicating probably lower quality of interfaces. The measurements carried out on voids as imaged by Photo Acoustic Microscopy did not indicate any signature of voids. Fig. 5 shows the spectra of a GeOI wafer at normal and oblique incidence using unpolarized light. The spectrum at normal incidence is characterized by strong Si-O-Si ASM at 1090cm-1 with a shoulder at 1170cm-1 and remaining bands at 460, 506, 562, 612, 740, 810 and 890cm-1. The most important feature of oblique incidence is the appearance and domination of the band at 1250cm-1. The strength of the bands at 506 and 810cm-1 are also increased with oblique incidence. There is no shift at the peak positions. Therefore, it can not be an effect of structural modulation, although there is some enhancement in transmission and fringes at around 900-2000cm-1. We rule out the possibility of attributing to a interface effects for the reason shown at the insert in Fig.5, where the normalized intensity of the band at 1250cm-1 is plotted versus the the effective oxide thickness(optical pathway). The extrapolation gives a value of about 0.46μm that is quite thick to be an interfacial layer. This behavior can be attributed to the effect of disorder induced mode coupling[2]. In thermally grown thin SiO2 on Si, coupled LO-TO frequency pairs are observed as peaks at around 1076-1256cm-1 in oblique incidence p-polarized absorption. Additionally two other LO-TO mode pairs were observed at 810-820cm-1 and 457-507cm-1[2] Therefore, the bands observed in GeOI at 506 and 810 are also related to LO-TO mode pairs. Ge bonding may be introducing an other component in light-matter interaction. In order to clarify the effect of Ge bonding, we measured the spectrum of a thermally grown $SiO_2$ on Ge. In fact, there is a similar oblique incidence effect, that is the enhancement of absorption shoulder at 1200cm-1. The strength of Si-O-Si band becomes strong with increased angle. This may suggest that Ge bonding also plays a role in disorder induced mode coupling. Ge is probably introducing an additional stress into $SiO_2$ layer. The other possibility can be based on arguments discussed in the previous section. n-type Ge is polarizing the light, thus leading to observed effects in GeOI. The birefringence is probably amplified by multiple reflections in SiO2 layer, n-Si and n-Ge wafers. Optical transmission nonlinearity depends on the first surface in GeOI. In general, enhancement in transparency was observed in both the cases, although they have different behavior. But, the enhancement and relaxation of transmission were found to be relatively slow process.

**Conclusion**

Infrared vibrational modes exhibit interesting behavior depending on oblique incidence and annealing. The appearance of new bands suggests the importance of longitudinal optic components of vibrations in Ge. The band at 666cm-1(probably VO-) is the most sensitive one to oblique incidence and annealing effects. Increased sub-gap absorption in highly Ga doped p-Ge is attributed to enhanced disorder. The Ge-Ge Raman phonon line broadening at 301cm-1 supports this observation. The bands at high energies may be due to deep levels or surface related species which need to be investigated further. Coupled LO-TO phonon modes at 1250cm-1 can be used to assess bonding in GeOI wafers. Observation of LO modes with unpolarized light may suggest the presence of a birefringent effect in Germanium. Optically created carriers lead to tranceparency in p-type, absorption in n-type Ge and side dependent nonlinearity in GeOI. These effects can be used to map carrier distribution in Ge and bonding interfaces in GeOI wafers.

**FIGURE CAPTIONS**

**Figure 1**     FTIR spectra of n-type Ge wafer at normal incidence, 40º unpolarized oblique incidence, undoped Ge and n-type germanium annealed under vacuum for 30 minutes.

**Figure 2**     FTIR spectra of p-type Ge waferwith normal and 80o oblique incidence and the spectrum of p-type Ge annealed at 400ºC under vacuum for 30minutes.  Inserted is an absorption dip at 808cm$^{-1}$ at low energies.

**Figure 3**     Transmission difference at 10.6μm in n-type(squares) and p-type(circles) Germanium wafer of 150μm thick as a function of UV-VIS pump power.

**Figure 4**     Transmission around the band gap of n- and p-Ge wafers.  Inserted is the Ge-Ge Raman phonon line in p-Ge wafer.

**Figure 5**     FTIR of GeOI wafer at different oblique incidence.  The band at 1250cm$^{-1}$ together with the bands at 810 and 611cm$^{-1}$ grows with increasing angle of incidence.

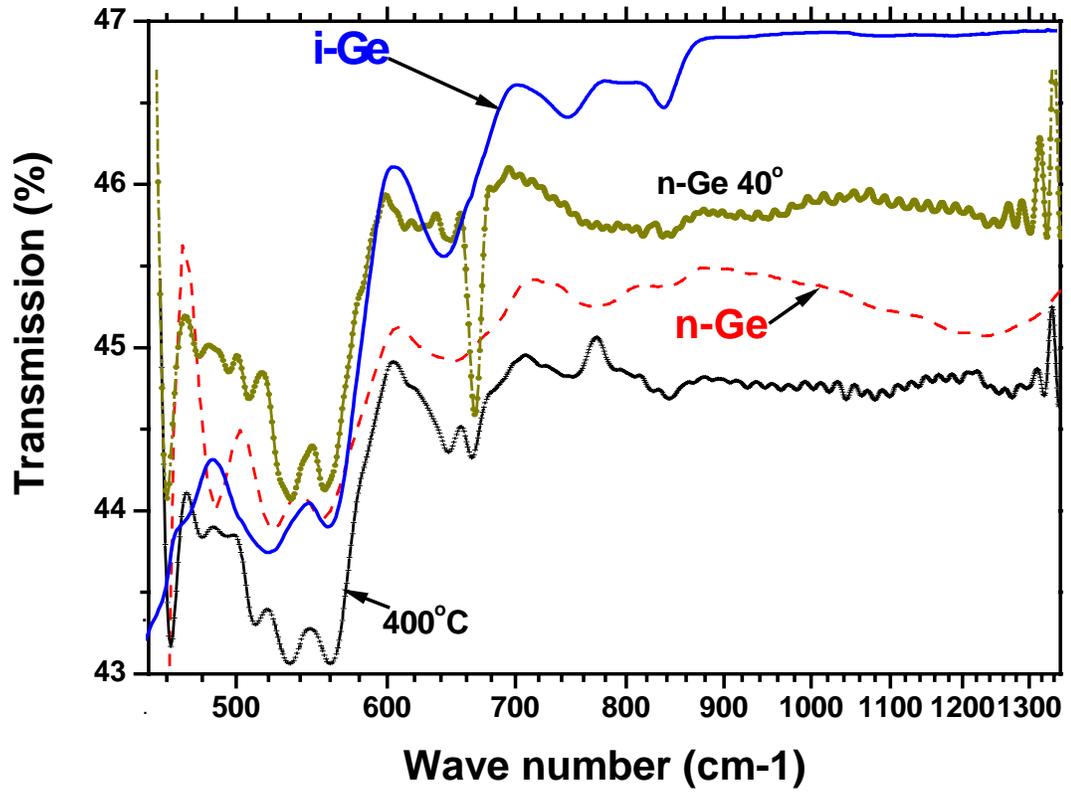

Figure 1

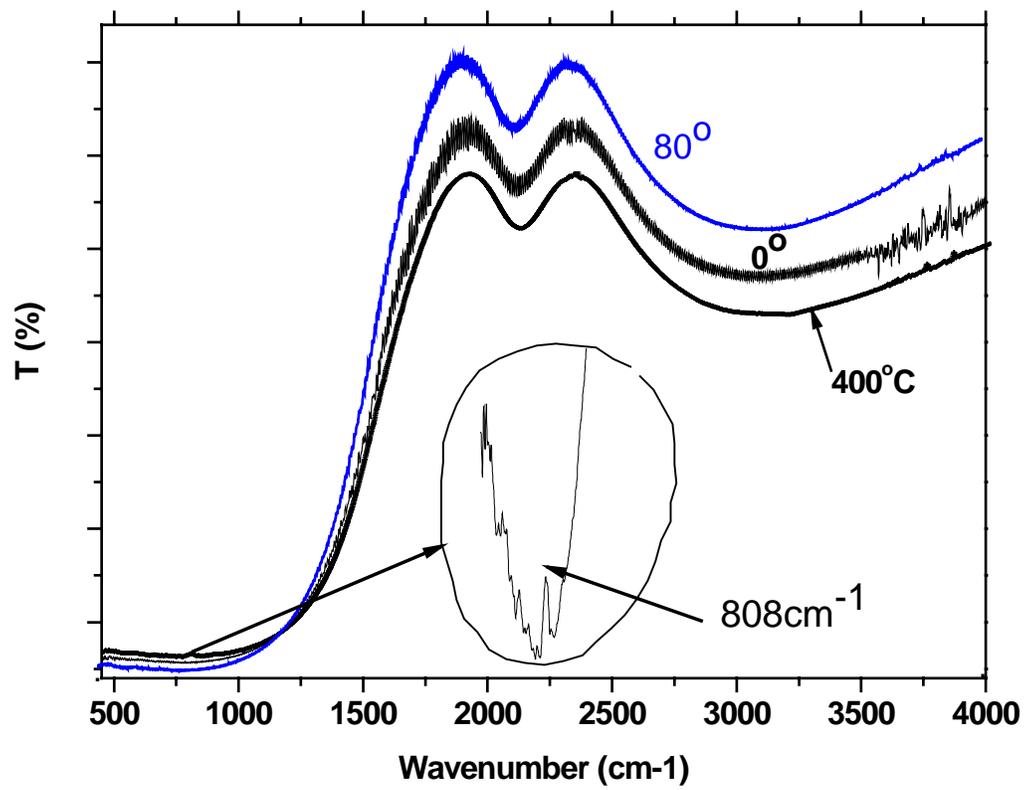

**Figure 2**

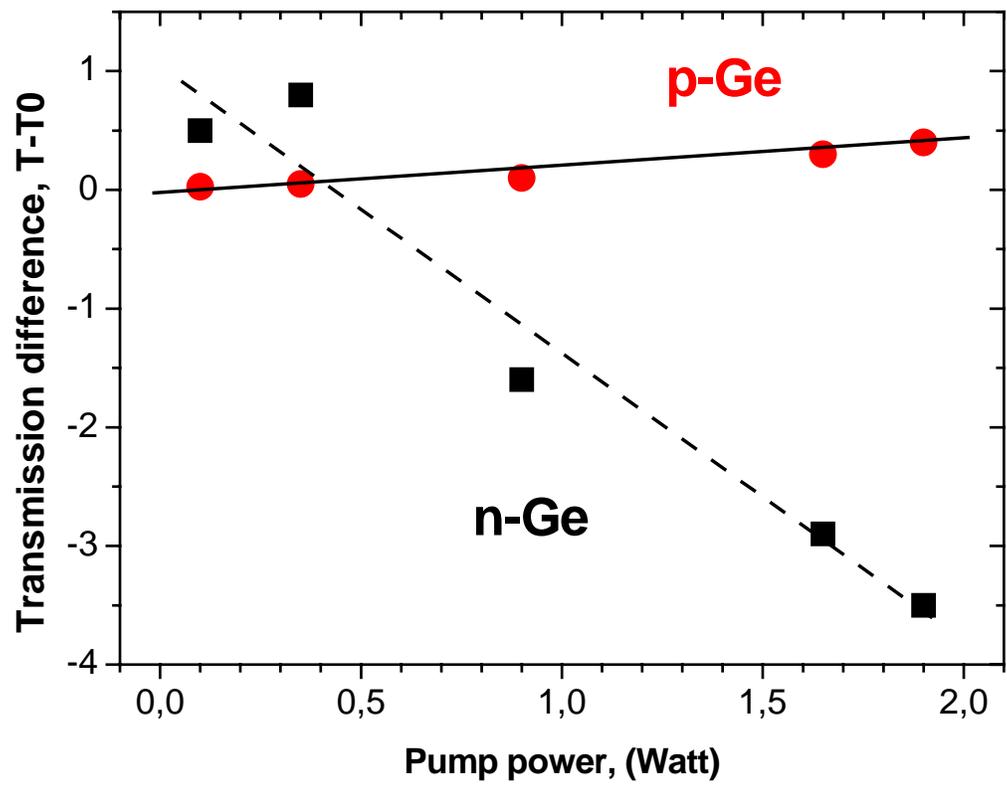

**Figure 3**

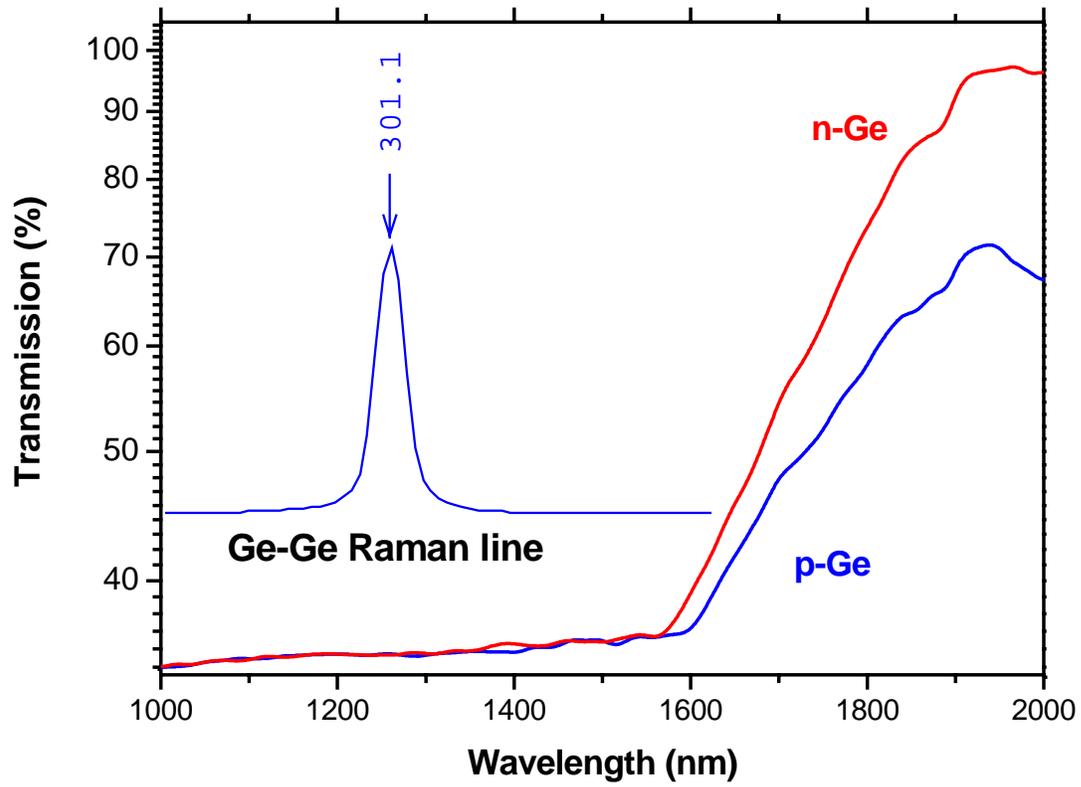

**Figure 4**

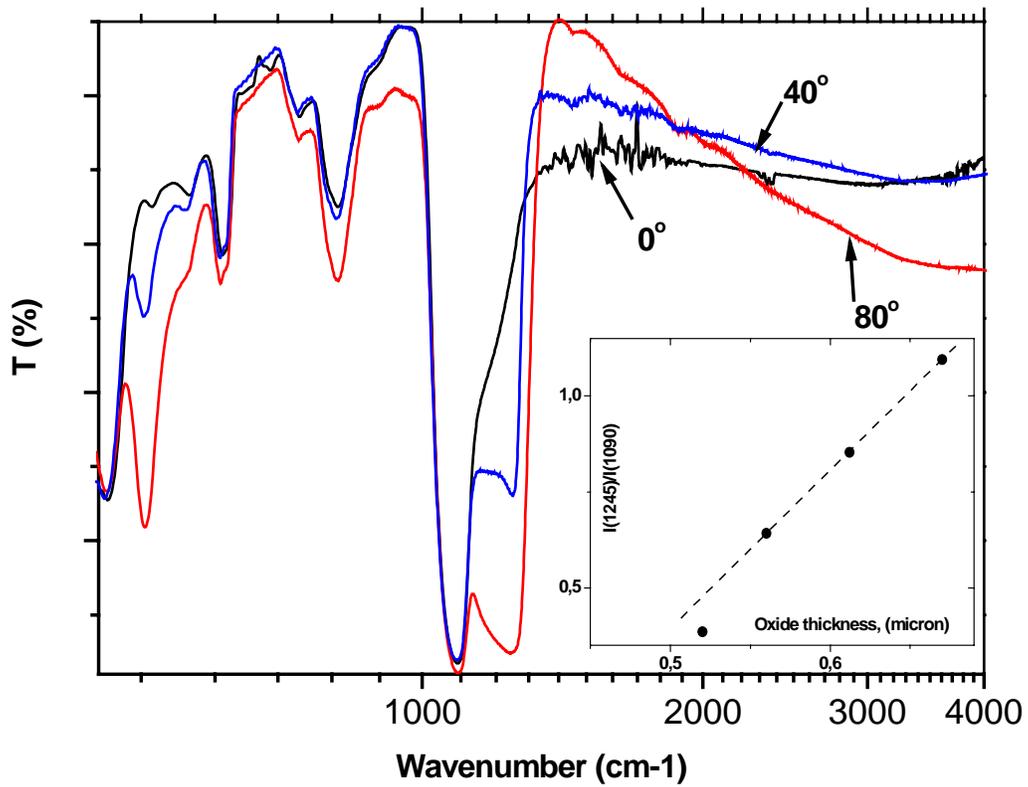

**Figure 5**